\documentstyle[11pt]{article}

\newcommand{\rf}[1]{(\ref{#1})}
\newcommand{\bea}{\begin{eqnarray}}
\newcommand{\eea}{\end{eqnarray}}

\newcommand{\g}{\gamma}

\renewcommand{\l}{\lambda}
\renewcommand{\b}{\beta}

\newcommand{\n}{\nu}
\newcommand{\m}{\mu}

\newcommand{\ep}{\varepsilon}

\newcommand{\del}{\delta}

\newcommand{\sg}{\sigma}

\newcommand{\oh}{\frac{1}{2}}
\newcommand{\oq}{\frac{1}{4}}

\newcommand{\ra}{\right\rangle}
\newcommand{\la}{\left\langle}
\newcommand{\prt}{\partial}

\newcommand{\cD}{{\cal D}}

\newcommand{\cL}{{\cal L}}
\newcommand{\cl}{{\pounds}}

\def\void{}
\def\labelmark{\marginpar{\small\labelname}}
\def\labelmark{}

\newenvironment{formula}[1]{\def\labelname{#1}
\ifx\void\labelname\def\junk{\begin{displaymath}}
\else\def\junk{\begin{equation}\label{\labelname}}\fi\junk}%
{\ifx\void\labelname\def\junk{\end{displaymath}}
\else\def\junk{\end{equation}}\fi\junk\labelmark\def\labelname{}}

{\ifx\void\labelname\def\junk{\end{array}\end{displaymath}}
\else\def\junk{\end{array}\right.\end{equation}}
\fi\junk\labelmark\def\labelname{}\def\junk{}
}

\newcommand{\beqj}{\begin{formula}}
\newcommand{\eeqj}{\end{formula}}
\newcommand{\beqv}{\begin{formula}{}}

\def\beq{\begin{equation}}
\def\eeq{\end{equation}}

\def\half{\frac{1}{2}}
\def\quart{\frac{1}{4}}

\def\gst{\gamma_{str}}

\def\mgap{\qquad \qquad}

\def\part{\langle 1 \rangle}
\def\ok{\frac{1}{k}}
\def\eok{{\epsilon^{1/k}}}
\def\deps{\Delta \epsilon}
\def\depsok{(\Delta \epsilon)^{1/k}}

\def\zvlog{\log \left( \frac{z-V}{z} \right) }
\def\pint{{\oint\limits_C \frac{dz}{2\pi i} \ }}
\def\iint{{\oint \frac{dz}{2\pi i} \ }}
\def\partv{\part_{_V}}
\def\dde{\frac{\partial}{\partial \epsilon}}

\def\gbar{\overline{g}}
\def\gpbar{\overline{g_p}}
\def\znbar{\overline{Z_N}(\gbar)}
\def\ubar{\overline{U}(M)}
\def\egbar{\overline{E_0}(\gbar)}
\def\tr{{\rm Tr}\ }
\def\punct{\left\langle P \right\rangle}
\def\pv{\punct_{_V}}

\def\back{\!\!\!\!}

\def\utw{{\widetilde{U}}}
\def\ubar{\overline{U}}
\def\llang{\left\langle}
\def\rrang{\right\rangle}

\def\liou{{_{\rm LV}}}

\begin{document}

\hfill    NBI-HE-97-08

\hfill February 1997

\begin{center}
\vspace{24pt}
{\Large \bf On the Connection Between 2d Topological Gravity\\
 and the Reduced Hermitian Matrix Model}

\vspace{36pt}

{\sl J. Ambj\o rn}\footnote{E-mail: ambjorn@nbi.dk}, 
{\sl M. G. Harris}\footnote{E-mail: Martin.Harris@nbi.dk}
 and {\sl M. Weis}\footnote{E-mail: weis@nbi.dk}  \\

\vspace{24pt}
 The Niels Bohr Institute,\\
Blegdamsvej 17, DK-2100 Copenhagen \O , Denmark.\\

\vspace{48pt}
  
\end{center}
\vspace{24pt}

\vfill

\begin{center}
{\bf Abstract}
\end{center}

\vspace{12pt}

\noindent
We discuss how concepts such as geodesic length and the volume of
space-time can appear in 2d topological gravity. We then construct a
detailed mapping between the reduced Hermitian matrix model and 
2d topological gravity at genus zero. This leads
to a complete solution of the counting problem for planar graphs with
vertices of even coordination number. The connection between
multi-critical matrix models and multi-critical topological gravity at
genus zero is studied in some detail.

\bigskip

\noindent
{\it PACS:} 04.60.-m

\noindent
{\it Keywords:} topological gravity; 2d quantum gravity; matrix model

\vfill

\newpage


\section{Introduction}

Despite substantial progress in our understanding of two-dimensional 
quantum gravity we are still left with two separate theories: 
two-dimensional topological gravity and Liouville theory.
Topological gravity is a theory where concepts such as geodesic distance
make no sense. It can be viewed as  a theory  on 
the moduli space of punctured Riemann surfaces, and it is defined 
in such a way that it allows us to calculate intersection 
indices related to the moduli spaces.
In Liouville theory geodesic distance plays a key r\^{o}le:
Hartle-Hawking wave functions are defined in terms of the geodesic 
length of the boundaries \cite{H-H,AmbJurMak90}
and it is possible to consider the quantum 
theory of universes of fixed space-time volume, as well as 
correlation functions depending on the geodesic distance 
between two punctures \cite{geodesics} (see also
references~\cite{japanese,kawai}
for a discussion of more general correlation functions).
Nevertheless, the two theories seem to be identical.
The various proofs of the equivalence have so far been rather 
indirect \cite{proofs,DVV91}.
Typically, generalized Schwinger-Dyson equations have been derived 
for the two theories and have been  shown to be identical by   
suitable identifications of variables. By such identifications
the concepts of length and area enter into topological gravity without any 
obvious interpretation. As an example let us mention the following 
situation: in topological gravity we can calculate the  
expectation values $\la \sg_n \ra$ of certain generalized puncture operators 
and we can define the generating function for this set of observables by 
\beqj{*1a}
\llang W(\ell) \rrang= \sum_n \frac{\ell^n}{n!} \la \sg_n \ra.
\eeqj 
In topological gravity $\ell$ has no interpretation, beyond that of an
indeterminate which defines the generating function. However,
$\llang W(\ell) \rrang$  agrees with the corresponding Hartle-Hawking 
wave functional calculated by the use of dynamical triangulations.
In Liouville theory  $\ell$ has from first principles 
an interpretation as the length of 
a macroscopic boundary of the two-dimensional universe. The same 
arguments apply to the total space-time volume $A$, which is 
defined from first principles if one uses Liouville theory, while 
it has no natural definition within the framework of topological gravity.

The rest of this paper is organized as follows: in
section~\ref{sec:two} we discuss how concepts such as the volume of
space-time and the length of boundaries can be introduced in
topological gravity. Since these concepts {\it are} natural in
topological quantum gravity it should be possible to make a detailed
mapping to Liouville theory. This is most conveniently done if we use
the dynamically triangulated version of Liouville theory.  In order to
define the mapping we introduce in section~\ref{sec:three} a contour
integral representation of the partition function of topological
gravity. A related integral representation for the Hermitian matrix
model allows us in section~\ref{sec:four} to perform a simple and
detailed identification between the correlation functions in
topological gravity and those in quantum gravity defined via dynamical
triangulations.  As a side result of this identification we use
topological gravity results to solve completely the counting problem
for planar graphs, whose vertices all have even coordination
numbers. In section~\ref{sec:hermmulti} we apply the mapping between
topological gravity and matrix models to the case of multi-critical
models. Section~\ref{sec:conc} contains our conclusions.

\section{Lagrangian Aspects of Topological Gravity}
\label{sec:two}

Various starting points are possible in the case of topological 
gravity. One approach is to start directly from 
geometric considerations involving the 
intersection indices on punctured Riemann surfaces~\cite{Wit90,DijWit}. 
This approach was completed and made mathematically rigorous by the seminal 
work of Kontsevich \cite{kontsevich}. From the remarkable properties of 
the Kontsevich integral it follows that the partition function of 
topological gravity is a tau-function for the Korteweg-de Vries 
equation. More precisely, the partition function is given by 
the following matrix integral
\beqj{*8}
Z_N(M) = \frac{\int d\phi \; \exp\Bigl\{ -\tr \Bigl( \frac{M \phi^2}{2}+
\frac{i \phi^3}{6} \Bigr)\Bigr\}}{\int d\phi \;
 \exp\Bigl\{ -\tr \Bigl( \frac{M \phi^2}{2}\Bigr)\Bigr\}},
\eeqj
where the integration is over $N\times N$ Hermitian matrices $\phi$.  
The matrix $M$ is assumed to be symmetric and positive definite, 
and  one can show that $Z_N(M)$ only depends on 
\beqj{*9}
t_k = - \ \frac{(2k-1)!!}{k!} \ \tr M^{-(2k+1)}, \mgap k \geq 0,
\eeqj
where by convention $(-1)!!=1$.
For  $N \to \infty$ the $u$-function of the KdV-equation is then
\beqj{*9a}
u(\{t_i\}) = \frac{\prt^2}{\prt t_0^2} \log Z(\{t_i\}).
\eeqj
One can write (for $N \to \infty$):
\beqj{*10}
\log Z (\{t_i\}) = \la \exp \sum_{j=0}^\infty t_j \sg_j \ra
= \sum_{n_0=0}^\infty \sum_{n_1=0}^\infty \cdots 
\la \sg_0^{n_0} \sg_1^{n_1} \cdots \ra \prod_{j=0}^\infty 
\left( \frac{t_j^{n_j}}{n_j !}\right) . \
\eeqj
This expression has a large $N$ expansion which is at the same 
time a genus expansion and it allows us in principle to calculate 
the numbers $ \la \sg_0^{n_0} \sg_1^{n_1} \cdots \ra_g$, which 
have the interpretation as intersection indices for each genus.

A number of proofs \cite{proofs}
exist of the equivalence between topological gravity 
formulated via \rf{*8} and Liouville gravity formulated via the 
use of dynamical tri\-an\-gu\-la\-ti\-ons.
As already mentioned the proofs of the equivalence between the theory 
defined by \rf{*8} and the theory of gravity defined via the matrix 
models corresponding to dynamical 
triangulations\footnote{We view here Liouville 
gravity and simplicial gravity as representing the same kind of 
quantum gravity, since observables referring directly to geometric objects
(Hartle-Hawking wave functionals etc.) 
can be defined in both theories, and since all calculations 
in the two theories agree so far.} are rather indirect and leave 
no hints of how to define observables associated with geodesic distances
in topological gravity. For this purpose it is more convenient to turn
to the original formulation \cite{LPW} of topological gravity, which 
originated from  Witten's general approach to topological field
theories~\cite{Witten-2}. 
Here the starting point for a topological theory is 
\beqj{*1}
\cL = 0.
\eeqj
The non-trivial content of \rf{*1} comes by defining the 
space of variables and performing a corresponding gauge fixing of the local 
symmetries. In the case of gravity one can choose as the field variable 
the metric $g_{\m\n}$. The Lagrangian \rf{*1} is invariant under the 
most general variation
\beqj{*q1}
\del g_{\m\n} = \kappa_{\m\n}.
\eeqj
It is common to factor out explicitly the invariance under 
diffeomorphisms. If $\xi_\m$ denotes an infinitesimal diffeomorphism and 
$r^\l_\m$ a general $2\times 2$ matrix, one can write
\beqj{*q2}
\del g_{\m\n} = \xi_{\m;\n} + \xi_{\n;\m} - (r^\l_\m g_{\l \n} +
r^\l_\n g_{\l \m}),
\eeqj
or, relating more closely to non-critical string theory, 
\beqj{*q3}
\del g_{\m\n}= \xi_{\m;\n} + \xi_{\n;\m} - \oh g_{\m\n} \xi^\l_\l +
g_{\m\n} \phi,
\eeqj
in decreasing order of generality (see \cite{Henneaux,CMR} for 
a discussion of the possible representations of \rf{*q1}). 
Gauge fixing according to either \rf{*q2}
or \rf{*q3} amounts to defining \rf{*1} to mean 
\beqj{*q4}
\cL_{_{\rm GF}} = s \cdot \Psi ,
\eeqj
where $s$ is the BRST operator for the gauge fixed symmetries. For  
both gauge fixings \rf{*q2} and \rf{*q3} the split into diffeomorphisms
and a residual shift is not unique, since there exist transformations 
different from the identity which leave $g_{\m\n}$ invariant. This 
leads to the well known ghosts for ghosts (super-ghosts) in $\cL_{_{GF}}$.
In the following we will use \rf{*q2} where the corresponding field  
content of the gauge fixed Lagrangian \rf{*q4} is the two-dimensional
metric $g_{\m\n}$, the gravitino field $\psi_{\m\n}$ (the ghost field
of the shift), the ghost field $c^\m$ of diffeomorphisms
and the super-ghost field $\g^\m$. 
In this representation, the action of the BRST operator $s$ is as
follows~\cite{LPW},
\begin{equation}
\begin{array}{ll}
s \cdot g_{\mu\nu} = \cl_c \ g_{\m\n} +\psi_{\m\n},~~~~~~~~ &
s \cdot c^\m = \oh \cl_c \ c^\m + \g^\m, \\
s \cdot \psi_{\m\n} \! = \cl_c \ \psi_{\m\n} -\cl_\g \ g_{\m\n}, & 
s \cdot \g^\m \! = \cl_c \ \g^\m,
\end{array}\label{brst}
\end{equation}
where $\cl_c$ and $\cl_\g$ denote the action of infinitesimal
diffeomorphisms with parameters $c^\m$ and $\g^\m$, and 
where $\psi_{\m\n}$ is related to 
the ghost $w^{\,\,\mu}_{\nu}$ for the matrix $r^{\,\,\mu}_{\nu}$ by
\begin{equation}
  \psi_{\mu\nu} = - w^{\,\,\gamma}_{\mu}g_{\gamma\nu} -
  w^{\,\,\gamma}_{\nu}g_{\gamma\mu}.
\end{equation}

One obtains the observables starting 
from the Euler 2-form~\cite{Becchi-al}
\beqj{*2}
\sg^{(2)} = \frac{1}{8\pi} \sqrt{g} R \; \ep_{\m\n} dx^\m \wedge dx^\n,
\eeqj
which gives rise  to a 1-form and a 0-form via the descent equations:
\beqj{*3}
s\cdot \sg^{(2)} = d \sg^{(1)},~~~~~s\cdot \sg^{(1)} = d \sg^{(0)},
 ~~~~~s\cdot \sg^{(0)}=0.
\eeqj
In this equation $s$ is the BRST operator and $d$ the 
exterior derivative. Explicitly one gets
\bea
\sg^{(0)} &=& \frac{\sqrt{g}\ep_{\m\n}}{4\pi}
\Bigl[ \oh c^\m c^\n R + c^\m D_\rho (\psi^{\n\rho}-g^{\n\rho}\psi^\sg_\sg)
+ D^\m\g^\n - \oq 
\psi^{\m}_\rho \psi^{\n\rho}\Bigr] \label{*4} ,\\
\sg^{(1)} &=& \frac{\sqrt{g}\ep_{\m\n}}{4\pi} \Bigl[ c^\n R +
 D_\rho (\psi^{\n\rho}-g^{\n\rho}\psi^\sg_\sg) \Bigr]dx^\m,\label{*5}
\eea
where $D_\m$ denotes the covariant derivative.
One can build an infinite set of observables from $\sg^{(0)}$ since 
the super-ghost $\g^\m$ is commutative:
\beqj{*6}
\sg^{(0)}_n \equiv \Bigl(\sg^{(0)} \Bigr)^n.
\eeqj

The expectation values of observables in topological gravity are 
now defined by functional integrals like
\beqj{*7}
\la \prod_k \sg_{n_k} (P_k) \ra \equiv \int \cD \Phi \; e^{-S[\Phi]}
\prod_k \sg_{n_k}(P_k),
\eeqj
where $\Phi$ denotes a suitable collection of fields (see below)
and the Lagrangian $\cL$ 
associated with the action $S[\Phi]$ is BRST exact.
The zero-forms $\sg_{n_k}$ sit at the 
points $P_k$ on the  world-sheet manifold and the super-gauge 
transformations should leave these points fixed. Correspondingly,
the fields $c$ and $\g$ must vanish at these points and this implies 
that we consider punctured Riemann surfaces. In this way the Riemann surfaces 
considered in the path integral become a function of the observables. 
This is in accordance with the interpretation of the ``observables''
$\tr \phi^{2l_1}\cdots  \tr \phi^{2l_k}$ in the Hermitian matrix model. 
The matrix integral corresponding to these observables is equivalent  
to a summation  over (generalized) triangulations with $k$ boundaries 
of length $2 l_1, \ldots , 2 l_k$, i.e.\ for 
each matrix observable we consider different triangulated surfaces. 
For fixed $l_1,\ldots,l_k$ the continuum lengths of the boundaries
go to zero and the loops become punctures. In later sections we will
make the correspondence between the $\sg_{n_k}$'s and the matrix model
observables precise.

In order to perform the functional integral \rf{*7} 
one usually chooses a background 
metric $g^{0}$ and background gravitino field $\psi^{0}$,
which depend on the moduli of the 
punctured Riemann surfaces, and after successive integration
over moduli all explicit reference to the points $P_k$ disappears~\cite{BI}.
In the following we will explicitly suppress all dependence
on moduli as they play no r\^{o}le in our arguments.
The background metric and background gravitino 
are introduced in a BRST exact way via 
Lagrange multiplier anti-ghosts 
 $b^{\mu\nu}$ and $\b_{\m\n}$, and their corresponding
Nakanishi-Lautrup fields $d^{\mu\nu}$ and $\del^{\m\n}$. The BRST
algebra extends trivially to these new fields 
\bea
 & s \cdot b^{\mu\nu} = d^{\mu\nu},\mgap   & s \cdot d^{\mu\nu} = 0,\nonumber\\
 & s \cdot \b^{\mu\nu} = \del^{\mu\nu}, \mgap & s \cdot \del^{\mu\nu} = 0.
\eea
The gauge fixed action is 
\begin{eqnarray}
  S_{_{\rm GF}} &=& \int_\Sigma d^2x\, \sqrt{g} \,\cL_{_{\rm GF}} =
\int_{\Sigma} d^{2}x\;\sqrt{g} \, s\cdot \Psi \nonumber\\
&=& \int_{\Sigma} d^{2}x\,\sqrt{g}\,
s\cdot \Bigl(b^{\mu\nu}(g_{\mu\nu}-g_{\mu\nu}^{0})+\b^{\m\n} 
({\psi}_{\m\n}-{\psi}^{0}_{\m\n})\Bigr) .
\end{eqnarray}

It is a remarkable fact that the functional integral \rf{*7}
allows the extraction of almost complete information about 
the intersection indices on the moduli space of punctured Riemann surfaces.
This is done by deriving Ward identities between various correlators and 
by showing that these Ward identities allow a recursive determination 
of expectation values of the $\sg_n$'s \cite{VV91,BI}.

Contrary to the Kontsevich integral formulation this 
definition of topological gravity operates with  space-time, but
offers {\it a priori} no concept 
of space-time distance.  Sometimes it is even said that there should be 
no such concept in quantum gravity, since by reparameterization
invariance, we only know if two points are separated or coincide and 
that this is the origin of the so-called contact terms.
This is in contrast to the approach to quantum gravity via Liouville 
theory, or, at the constructive level, via dynamical triangulations.
To be more explicit we can in Liouville theory
define the partition function for a fixed space-time volume by 
\beqj{*11a}
Z_\liou (A) = \int \cD [g] \;\del \left(\int d^2 \xi \sqrt{g} - A\right) ,
\eeqj
and a  Hartle-Hawking 
wave functional corresponding to a fixed boundary length $\ell$ and fixed 
space-time volume $A$ by:
\beqj{*11}
W_\liou(\ell,A) = 
\int \cD [g] \; \del \left(\int d^2 \xi \sqrt{g} - A \right)\;\del
\left( \oint ds -\ell \right).
\eeqj
Similarly we can define the partition function of  quantum universes  
with two marked points separated by a geodesic distance $R$ and with 
space-time volume $A$ by
\beqj{*12}
G_\liou(R,A) = \int \cD[g] \; \del \left(\int d^2 \xi \sqrt{g} - A \right) \, 
\int d^2 \xi \sqrt{g} \int d^2 \xi' \sqrt{g} \; \del 
\left(D_g (\xi,\xi')-R \right), 
\eeqj
where $D_g (\xi,\xi')$ is the geodesic distance between $\xi$ and $\xi'$
with respect to the metric $g$. 
In these equations
the functional integration is over all equivalence classes of metrics.
In a certain sense  $Z_\liou(A)$, $W_\liou(\ell, A)$ and $G_\liou(R,A)$ 
are ``more topological''
than topological gravity itself in the spirit of being a functional 
integral with  $\cL=0$, since 
no action is ever used in \rf{*11a}--\rf{*12}. However, at the 
same time they have explicit reference to metrical properties of the 
quantum space-time. In fact $G_\liou(R,A)$ is a perfect probe of the 
fractal structure of space-time~\cite{geodesics}
\footnote{Strictly speaking it has not 
been possible to calculate $W_\liou(\ell,A)$ and $G_\liou(R,A)$ entirely 
within the continuum framework of Liouville theory. However,
they can be calculated using  dynamical triangulations
\cite{H-H,AmbJurMak90,geodesics}.}.

We can also construct a partition function for topological gravity which
corresponds to a fixed space-time volume. To do this we introduce the
appropriate delta functions as a global, BRST exact term in the action,
\beqj{*q11}
Z[A_0] = \int \cD \Phi d\chi d\m \; 
\exp\left\{-S_{_{\rm GF}}-s\cdot \left[ \chi \left(
\int_\Sigma  \sqrt{g}-A_0 \right) \right] \right\}, 
\eeqj
by use of Lagrange multipliers forming a constant
anti-ghost
multiplet $(\chi,\m)$, which transforms like
\beqj{*q12}
s\cdot \chi = \m,\mgap s \cdot \m =0.
\eeqj  
In contrast to equation \rf{*11a} this also includes an additional
delta function
in the gravitino field, induced by supersymmetry. We must
demand that the area of the surface is compatible with the chosen
background metric, that is, $A_0 = \int_\Sigma \sqrt{g_0}$.
By Laplace transformation of \rf{*q11}
one finds a new topological action with a BRST exact ``cosmological term'',
\beqj{*q11b}
Z[\mu] = \int \cD \Phi d\chi  \; 
\exp\left\{ -S_{_{\rm GF}}-s\cdot \left(\chi
\int_\Sigma \sqrt{g} \right)\right\}. 
\eeqj
If the surface $\Sigma$ has  boundaries it is also possible 
to introduce explicitly a ``boundary cosmological term'' living
on the boundary of $\Sigma$ \cite{montano}, and 
by an inverse Laplace transformation like \rf{*q11} one can fix the 
length of the boundary and in this way obtain a topological gravity 
equivalent of \rf{*11}\footnote{We believe that it is possible to 
define a topological gravity partition function corresponding 
to \rf{*12}, too, but since we do not know how to calculate 
the corresponding partition function within the framework 
of topological gravity as a function of geodesic 
distance, we have not pursued this question further.}.

The above considerations  show that one can indeed define 
topological gravity on a restricted moduli space of metrics where 
for instance the area $A$ has a fixed value. In this way there is no conflict 
between the  appearance of observables that refer 
to global metrical properties, such as the total area or the 
total length of a boundary, and the concept of topological gravity.
We now proceed to establish a precise mapping between the 
operators $\sg_n$ as they appear in topological gravity and the 
corresponding operators in the language of dynamical triangulations
or matrix models. 

\section{Topological Gravity at Genus Zero}
\label{sec:three}

For the purpose of establishing this required map between topological 
gravity and matrix models we will consider genus zero surfaces only.
Further it is simplest to start from the formulation of 
topological gravity found in \cite{Wit90}.

\subsection{Contour Integral Representation}

In reference~\cite{Wit90} Witten considers 2d topological gravity with
a perturbed action. The theory contains a set of basic operators,
$\{\sigma_n\}$, where $n$ is an integer with $n\geq 0$, and the
perturbation is parameterized by a set of coupling constants, $\{t_n\}$.
The unperturbed
correlation functions will be written with the subscript $V\!\! =\!\! 0$; below
we introduce a potential $V(z)$, which depends on the perturbed
action, and $V\!\! =\!\! 0$ corresponds to the unperturbed case. Then the
partition function, $\partv$, for the perturbed theory at genus $g$ is
given in terms of the unperturbed correlation functions for the same
genus, 
\begin{eqnarray}
\label{eq:part}
\partv & \equiv & \llang \exp \left[ \sum_{j=0}^{\infty} t_j \sigma_j
\right] \rrang_{\scriptscriptstyle  \! \! V=0} \\
&=& \sum_{n_0=0}^{\infty} \sum_{n_1=0}^{\infty} \cdots \left( 
\prod_{j=0}^\infty
 \frac{t_j^{n_j}}{n_j!} \right) \llang (\sigma_0)^{n_0}
(\sigma_1)^{n_1} (\sigma_2)^{n_2} \cdots
\rrang_{\scriptscriptstyle {V=0}} .
\end{eqnarray}

For genus zero the unperturbed correlation functions can be easily
calculated from the puncture and dilaton equations~\cite{Dij91}.
With the normalization of $\sigma_n$ used in reference~\cite{Wit90},
we have,
\beq
\llang \prod_{i=1}^{s} \sigma_{d_i} \rrang_{_{\!\! V=0}} = 
\left(\sum_{i=1}^{s} d_i \right) ! \ \ \delta\left( 3+ \sum_{i=1}^{s}
(d_i - 1) \right),
\eeq
where $\delta(n)\equiv \delta_{n,0}$.
The correlation function of a product of observables for unperturbed
topological gravity vanishes unless the total ghost number of the
observables equals the dimension of the moduli space; the delta
function reflects this fact.
At higher genus the unperturbed correlation
functions are much more complicated.
The rest of
this paper is almost exclusively concerned with genus zero
surfaces and all the correlation functions given are for genus zero, unless
it is stated otherwise.

Now we will define a potential $V(z)$ for the theory given in equation
(\ref{eq:part}),
\beq
V(z)= \sum_{j=0}^{\infty} t_j z^j ,
\eeq
where we have introduced a new complex variable $z$. 
We will write $t_0$ as $\epsilon$ and the puncture operator
$\sigma_0$ as $P$, when it is convenient to do so.
Then we can rewrite the partition function for the perturbed theory at
genus zero as a contour integral. We have
\begin{eqnarray}
\label{eq:partseries}
\partv &=& \sum_{n_0=0}^{\infty} \sum_{n_1=0}^{\infty} \cdots \left(
\prod_{j=0}^\infty  \frac{t_j^{n_j}}{n_j!} \right)
\left( \sum_{k=0}^{\infty} k n_k \right) ! \ \ \delta \left( 3+
\sum_{k=0}^{\infty} (k-1) n_k \right) \\
&=& \left[ \left(1-\frac{\partial}{\partial \lambda} \right)^{-1} 
\sum_{n_0=0}^{\infty} \sum_{n_1=0}^{\infty} \cdots
\prod_{j=0}^\infty \left( \frac{t_j^{n_j}}{n_j!}  \ z^{(j-1)n_j}
\lambda^{j n_j}\right) \right]_{\lambda=0 \atop {\rm coeff.\ of}\ z^{-3}},
\end{eqnarray}
where
\beq
\left(1-\frac{\partial}{\partial \lambda} \right)^{-1} \equiv
\sum_{n=0}^{\infty} \left(\frac{\partial}{\partial \lambda} \right)^n.
\eeq
The derivatives with respect to $\lambda$, give the factor of $(\sum k
n_k)!$ and keeping only the coefficient of $z^{-3}$ enforces the delta
function. Thus,
\beq
\partv = \left[ \left(1-\frac{\partial}{\partial \lambda}
\right)^{-1} \iint z^2 \exp \left({\frac{1}{z} V(\lambda
z)}\right) \right]_{\lambda=0},
\eeq
where we are integrating around some suitable contour encircling the
origin. Changing variables $z \to z/\lambda$ gives
\beq
\partv = \left[ \left(1-\frac{\partial}{\partial \lambda} \right)^{-1}
\iint \frac{z^2}{\lambda^3} \exp\left( {\frac{\lambda}{z}
V(z)} \right) \right]_{\lambda=0}.
\eeq
Expanding the exponential in powers of $\lambda$ and performing the
differentiation with respect to $\lambda$ gives
\beq
\partv = \sum_{r=3}^{\infty} \iint z^2 \left(
\frac{V(z)}{z} \right)^r \frac{1}{r(r-1)(r-2)} .
\eeq
It is convenient to rescale $V$ by a factor of $\beta$; then after
$V(z) \to \beta V(z)$ we have
\beq 
\frac{\partial^3}{\partial \beta^3} \part_{_{\beta V}} = \iint
\left(\frac{V^3}{z}\right) \frac{1}{1- {\beta V}/{z}}.
\eeq
Integrating with respect to $\beta$ three times and then
setting $\beta=1$ gives
\beq
\label{eq:contour}
\partv = -\half \ \pint (z-V)^2 \log
\left(\frac{z-V}{z} \right) .
\eeq
This formula, which gives the genus zero partition function for 2d
topological gravity with an arbitrary polynomial perturbation $V(z)$,
will be used extensively throughout the rest of the paper.
For a suitable choice of the parameters $\{t_n\}$, 
the integrand has a cut on
the positive real axis from $z=0$ to some point, which we will call
$z=u$. This point is defined by the equation $u=V(u)$, that is, 
\beq
u= \epsilon + \sum_{k=1}^{\infty} t_k u^k,
\eeq
which is just the well-known string equation for genus zero~\cite{DijWit}. 
The contour $C$ encircles
this cut, across which the integrand has a discontinuity of $-2\pi i$.
It will be convenient to assume that $\epsilon \geq 0$.

\subsection{Multi-Critical Models}
\label{sec:multitop}
In this section we will calculate various correlation functions for
the multi-critical models. The simplest choice of potential for the
$k$-th multi-critical model is $V= \epsilon + z - z^k$, that is $t_0 \neq
0$, $t_1 =1$ and $t_k=-1$. Note however that this choice of potential is
not unique and in section~\ref{sec:hermmulti} we will show that the
usual Hermitian one-matrix multi-critical models correspond to a different
choice of $V(z)$. The scaling behaviour for the $k$-th multi-critical
model is however independent of the particular choice of $V(z)$.

 From (\ref{eq:contour}) we have that
\beq
\part = -\half \ \pint \left( z^k - \epsilon
\right)^2 \log \left(\frac{z^k - \epsilon}{z} \right) ,
\eeq
for the $k$-th multi-critical model. There is a cut on the real axis in
the interval $\left[0, \epsilon^{1/k}\right]$, which is encircled by
the contour $C$. Thus,
\beq
\label{eq:partscale}
\part = \half \int_{0}^{\eok} \!\!\!\! dx \left( x^k - \epsilon \right)^2 =
\frac{\epsilon^{2+\ok}}{(2+\ok)(1+\ok)},
\eeq
which agrees with the result in reference~\cite{Wit90}.
Note that the string susceptibility, $\gst$, can be defined by $\part
\sim \epsilon^{2-\gst}$. Thus we see that $\gst=-\ok$ for the $k$-th
multi-critical model.

The one-point correlation function can be obtained by differentiating
(\ref{eq:contour}) with respect to $t_{l_1}$. This gives for a general
potential $V(z)$,
\beq
\label{eq:onepoint}
\llang \sigma_{l_1} \rrang_{_V} = \pint z^{l_1}
\left(z-V \right) \zvlog  .
\eeq
Specializing to the $k$-th multi-critical model gives
\beq
\llang \sigma_{l_1} \rrang = - \int_0^{\eok} dx \ x^{l_1}
\left( x^k - \epsilon \right) = \frac{\epsilon^{1+(l_1 +1)/k}}{(l_1
+1)(1+(l_1 +1)/k)}.
\eeq
Similarly the two-point function is
\beq
\label{eq:twopoint}
\llang \sigma_{l_1} \sigma_{l_2} \rrang_{_V} =
- \pint z^{l_1 +l_2} \zvlog = \int_0^u dx \ x^{l_1+l_2} =
\frac{u^{l_1+l_2+1}}{(l_1+l_2+1)} .
\eeq
Hence for the multi-critical model,
\beq
\llang \sigma_{l_1} \sigma_{l_2} \rrang 
= \frac{\epsilon^{(l_1+l_2+1)/k}}{(l_1+l_2+1)}.
\eeq
For the general potential we also have that
$u=\langle PP \rangle_{_V}$, a relationship which will prove useful later.
The three-point function is
\begin{eqnarray}
\llang \sigma_{l_1} \sigma_{l_2} \sigma_{l_3} \rrang_{_V} &=&
\pint \frac{z^{l_1+l_2+l_3}}{z-V} \\
&=& \frac{\partial}{\partial \epsilon} \left[ - \pint
z^{l_1+l_2+l_3} \zvlog \right]
\end{eqnarray}
and in general,
\beq
\label{eq:genfunc}
\llang \sigma_{l_1} \sigma_{l_2} \cdots \sigma_{l_s}
\rrang_{_V} = \left(\dde\right)^{s-2} \left[ - \pint
z^L \zvlog \right] = \left(\dde \right)^{s-2} \left[
\frac{u^{L+1}}{(L+1)} \right] ,
\eeq
where $L \equiv \sum\limits_{i=1}^{s} l_i$.
For the $k$-th multi-critical model, we put $u=\eok$ in this formula.
This result
agrees with the calculation in references~\cite{Wit90,DijWit} for the
multi-critical models in topological gravity and also the similar
result for the correlation functions in the multi-critical
Hermitian matrix model~\cite{GroMig}; this latter model is studied in
section~\ref{sec:hermmulti}.

\subsection{Loop Operators}
\label{sec:loop}
In this section we define an operator $W(l)$, which we will call
a loop operator, by analogy with similar observables defined in matrix
models. The form of (\ref{eq:onepoint}) suggests that we should define
the loop operator as
\beq
W(l) = \sum_{n=0}^{\infty} \frac{l^n}{n!} \sigma_n .
\eeq
Let us now calculate expectation values of products of $W(l)$ for a
general potential and for the multi-critical models. The results in the
previous section immediately give for the potential $V(z)$,
\beq
\llang W(l_1) \rrang_{_V} = \pint e^{l_1 z} (z-V)
\zvlog .
\eeq
For the $k$-th multi-critical model this gives
\beq
\llang W(l_1) \rrang = - \int_{0}^{\eok} \back dx \ e^{l_1 x}
\left( x^k - \epsilon \right) = \left[ \epsilon -
\left(\frac{\partial}{\partial l_1}\right)^k \right] \left(
\frac{e^{l_1 \eok}-1}{l_1} \right) .
\eeq
For $s \ge 2$,
\begin{eqnarray}
\llang \prod_{i=1}^{s} W(l_i) \rrang_{_{\!\! V}} &=& \left(\dde
\right)^{s-2}  \left[ - \pint \exp \left( z \sum_{i=1}^{s} l_i
\right) \zvlog \right] \\
& = &
\left(\dde\right)^{s-2} \left[\frac{e^{L u} -1}{L} \right],
\end{eqnarray}
where $L=\sum\limits_{i=1}^{s} l_i$ as before.
The $k$-th multi-critical model is given by setting $u=\eok$ in this
last equation.
The relationship between these loop operators and the corresponding
correlators which occur in matrix models is discussed in
section~\ref{sec:mmloop}.

\subsection{The Dilaton and Puncture Equations}
The dilaton equation can be derived from (\ref{eq:contour}) in a
simple fashion; we have
\beq
\sum_{k=0}^{\infty} t_k \frac{\partial}{\partial t_k} \partv = \pint V
(z-V) \zvlog 
\eeq
and this gives us the dilaton equation,
\beq
\left[ \frac{\partial}{\partial t_1} - \sum_{k=0}^{\infty} t_k
\frac{\partial}{\partial t_k}  +2 \right] \partv =0.
\eeq
We will also derive the puncture equation. We have
\beq
\left[ \dde - \sum_{k=0}^{\infty} (k+1) t_{k+1}
\frac{\partial}{\partial t_k} \right] \partv = \pint \left[
\frac{d}{dz} (z-V) \right] (z-V) \zvlog .
\eeq
After integrating by parts the right-hand side can be evaluated to
give the puncture equation,
\beq
\left[ \dde - \sum_{k=0}^{\infty} (k+1) t_{k+1}
\frac{\partial}{\partial t_k} \right] \partv = \half \epsilon^2.
\eeq

Many of the results in these sections have been derived by different
methods in a number of papers~\cite{Wit90,DijWit,VV91}. So far
the contour integral representation of 2d topological gravity, at
genus zero, has allowed us to reproduce the formulae for various
correlation functions in a very simple fashion. However, the real
power of the method will become apparent when we consider the
Hermitian one-matrix model in the next section.

\section{Hermitian Matrix Model}
\label{sec:four}
\subsection{Contour Integral Representation}
\label{sec:firstmodel}
Consider an Hermitian matrix model with an even potential; this is
sometimes referred to as the reduced Hermitian matrix model. The
partition function is defined as
\beq
\label{eq:defmat}
\znbar = \int {\cal D}M \ \exp \left[ -  \tr \ubar \right],
\eeq
where $M$ is an $N \times N$ Hermitian matrix and the potential
$\ubar$ is given by
\beq
\label{eq:defpot}
\ubar(M) = \half M^2 + \sum_{p=2}^{\infty} \frac{\overline{g_p}}{N^{p-1}}
M^{2p} .
\eeq
Note that we are writing the variables with bars on them in order to
differentiate them from the rescaled variables, which will be
introduced in the next section. The free energy at genus zero is given by
\beq
\egbar = \lim_{N \to \infty} - \frac{1}{N^2} \log \left(\frac{\znbar}{
\overline{Z_N}(0)}\right).
\eeq

The problem of calculating $\egbar$ has been solved using the
technique of orthogonal polynomials and in reference~\cite{Bessis} it
is shown that
\beq
\label{eq:egbarw}
\egbar = \int_{0}^{a^2} dr \ w'(r) \left[ 1 -w(r) \right] \log \left(
\frac{w(r)}{r} \right) ,
\eeq
where
\beq
w(r) = r + \sum_{p=2}^{\infty} \overline{g_p} \frac{(2p)!}{p!(p-1)!} r^p 
\eeq
and $a^2$ is the solution of
\beq
1=w(a^2), \mgap a^2>0 .
\eeq
Integrating by parts gives
\beq
\label{eq:ebarint}
\egbar = \half \int_{0}^{a^2} dr \ (1-w)^2 \frac{d}{dr} \left[
\log\left( \frac{w(r)}{r}\right)\right].
\eeq
We would like to rewrite this equation as a contour integral in such a
way as to make the connection with 2d topological gravity clear. First
let us change variables from the set of parameters $\{\overline{g_p}\}$ to
the set $\{t_p\}$ by defining,
\beq
\label{eq:relate}
\frac{\overline{g_p}(2p)!}{p!(p-1)!} = - \frac{t_p
\epsilon^{p-1}}{(1-t_1)^p} \mgap {\rm for \ } p \ge 2.
\eeq
Note that the set of parameters $\{t_p\}$ has two more variables than the
original set $\{\overline{g_p}\}$, 
namely $\epsilon$ and $t_1$; this gives us some
freedom in the choice of these last two variables. When it is
convenient we will write $\epsilon$ as $t_0$. In terms of these new
parameters and using the variable $z$ in place of $r$, $w$ is given by
\beq
\label{eq:wtov}
w(z) -1 = \frac{1}{\epsilon} \left[ z' - V(z') \right],
\eeq
where
\beq
z' = \frac{\epsilon z}{1-t_1}
\eeq
and $V(z)$ is defined as for topological gravity,
\beq
\label{eq:pot}
V(z) = \sum_{p=0}^{\infty} t_p z^p.
\eeq
The integral in (\ref{eq:ebarint}) is along the real axis from zero to
$a^2$, where $a^2$ is given by the condition $w(a^2)=1$. This
corresponds to the range $z'=0$ to $z'=u$, where $u$ is defined by
$u=V(u)$. This is precisely where the cut occurs on the real axis for
topological gravity with the potential $V$ and suggests that we should
rewrite (\ref{eq:ebarint}) as
\beq
\egbar =  - \half \ \iint \left[1-w(z) \right]^2
\left[\frac{d}{dz} \log \left(\frac{w(z)}{z}\right) \right] \log
\left( \frac{w(z)-1}{z} \right) .
\eeq
The contour encircles the cut from $z=0$ to $z=a^2$, which has been
introduced by adding the final logarithm in the integrand. Having
rewritten $\overline{E_0}$ as a contour integral all that remains is
to eliminate $w(z)$ in favour of $V(z)$, using (\ref{eq:wtov}),
\beq
\overline{E_0}= - \frac{1}{2 \epsilon^2} \pint
(z-V)^2 \left[ \frac{d}{dz} \log \left( \frac{z-V+\epsilon}{z} \right)
\right] \zvlog ,
\eeq
where the contour $C$ is around the cut $[0,u]$ on the real axis. This
gives the free energy at genus zero for the reduced Hermitian
matrix model, and clearly the result is very similar to the genus zero
partition function of topological gravity, equation
(\ref{eq:contour}). Equation (\ref{eq:relate}) relates the parameters in
the two models.

\subsection{Rescaled Matrix Model}
\label{sec:rescale}
The subsequent analysis will be simpler if we rescale the matrix in
the matrix model, removing all the factors of $N^{1-p}$ and
introducing a parameter in front of the quadratic term in the
potential. We introduce a matrix $\Phi = M (1-t_1)^{-\half}
N^{-\half}$, set $g_p=\overline{g_p} (1-t_1)^p$ for $p \ge 2$ and
define $g_1 = - \half t_1$.
Then after substituting into equations (\ref{eq:defmat}) and
(\ref{eq:defpot}) we have
\beq
\label{eq:zred}
Z_N(g) = \int {\cal D} \Phi \ \exp \left[ - N \tr U(\Phi) \right],
\eeq
where
\beq
\label{eq:upot}
U(\Phi) = \half \Phi^2 + \sum_{p=1}^{\infty} g_p \Phi^{2p}
\eeq
and the genus zero free energy is
\beq
\label{eq:freeenergy}
E_0(g) = \lim_{N \to \infty} - \frac{1}{N^2} \log \left(
\frac{Z_N(g)}{Z_N(0)} \right) .
\eeq
The contour integral representation is now,
\beq
E_0 = \half \log(1-t_1) - \frac{1}{2 \epsilon^2} 
\pint (z-V)^2 \left[ \frac{d}{dz} \log \left( 
\frac{z-V+\epsilon}{z} \right) \right] \zvlog ,
\eeq
with $V(z)$ defined by (\ref{eq:pot}) as before. The relationship
between the two sets of parameters is 
\beq
\label{eq:newrelate}
\frac{g_p(2p)!}{p!(p-1)!} = - {t_p
\epsilon^{p-1}} \mgap {\rm for \ } p \ge 1.
\eeq
For a given $U(\Phi)$ we may freely choose $\epsilon$ (with $\epsilon
\neq 0$) and then all the other $t_p$ are determined.

The relationship between the reduced Hermitian matrix model and
topological gravity is simpler if we differentiate $E_0$ with
respect to $t_l$. For $l\ge 1$,
\beq
\frac{\partial E_0}{\partial t_l} = 
\frac{- \delta_{l,1}}{2(1-t_1)} - \frac{1}{2 \epsilon^2} 
\pint \left[ 2 z^{l-1} (z-V) \zvlog - \frac{z^{l-1}
(z-V)^2}{z-V+\epsilon}  \right] ,
\eeq
which we gain after differentiating under the integral sign,
integrating by parts and dropping analytic terms in the integrand.
However the last term in the integrand cancels with the first term on
the right-hand side, giving
\beq
\label{eq:mainresult}
\frac{\partial E_0}{\partial t_l} = - \frac{1}{\epsilon^2}
\llang \sigma_{l-1} \rrang_{_V}
\eeq
and hence for $l \ge 1$,
\beq
\label{eq:oneptrelate}
\llang \sigma_{l-1} \rrang_{_V} = \lim_{N \to \infty}
\frac{\epsilon^{l+1}}{N} \ \frac{l!(l-1)!}{(2l)!} \llang \tr
\Phi^{2l} \rrang .
\eeq
This equation relates, at genus zero, the one-point correlation
functions in topological gravity with correlation functions in the
corresponding reduced Hermitian matrix model. Setting $l=1$ gives
\beq
\llang P \rrang_{_V} = \lim_{N \to \infty} \frac{1}{N}
\frac{\epsilon^2}{2} \llang \tr \Phi^{2} \rrang ,
\eeq
which shows that the correlation function for genus zero surfaces with
a single puncture in topological gravity, is proportional to the
partition function for triangulated surfaces with a marked link, with
a suitable identification of the coupling constants in the two
theories, that is, equation (\ref{eq:newrelate}).

In a similar fashion, one can show by differentiating $E_0$ with
respect to $\epsilon$ that,
\beq
\label{eq:dedeps}
\epsilon^3 \frac{\partial E_0}{\partial \epsilon} = -
\sum_{l=1}^{\infty} l \ t_{l+1} \llang \sigma_l \rrang_{_V} .
\eeq

Equation (\ref{eq:oneptrelate}) can be generalized to multi-point
correlators. Let us define a set of operators, $\{\Sigma_l\}$, in the matrix
model as follows,
\beq
\Sigma_l = N \epsilon^{l-1} \ \frac{l! (l-1)!}{(2l)!} \ \tr \! \! \left( 
\Phi^{2l} \right) \mgap {\rm for \ } l \ge 1 .
\eeq
Then we have, by repeatedly differentiating (\ref{eq:oneptrelate})
with respect to $t_{l_i}$,
\beq
\llang \sigma_{l_1-1} \prod_{i=2}^{s} \sigma_{l_i}
\rrang_{\! \! _V} = \lim_{N \to \infty} \left( \frac{\epsilon}{N}
\right)^2 \llang \prod_{i=1}^s \Sigma_{l_i} \rrang_{\rm
\! \! conn.} \mgap {\rm for \ } l_i \ge 1 ,
\eeq
where $\llang \cdots \rrang_{\rm conn.}$ denotes the
connected part of the expectation value in the matrix model; that is,
\beq
\llang \Sigma_{l_1} \Sigma_{l_2} \rrang_{\rm conn.}
\equiv \llang \Sigma_{l_1} \Sigma_{l_2} \rrang -
\llang \Sigma_{l_1}  \rrang
\llang \Sigma_{l_2}  \rrang
\eeq
and so on.

Thus it is clear that the operator $\sigma_l$ in topological gravity
corresponds to the operator $\Sigma_l$ in the matrix model, although
the relationship is slightly different for the first operator. Note that
this identification differs somewhat from the usual one in the
literature~\cite{Wit90, DVV91, KleWil, Distler},
as many papers confuse $\sigma_l$ with an
operator $O_l$, which is related to the $l$-th
multi-critical model. This point will be discussed further in
section~\ref{sec:hermmulti}.

\subsection{Power Series Expansion}
In this section the connection between topological gravity and the
Hermitian matrix model is used to write the genus zero free energy of
the matrix model as a power series. It is simplest to begin with the
version of the matrix model defined in
section~\ref{sec:firstmodel}. Using the relationship,
\beq
\label{eq:barnobar}
\overline{E_0} = E_0 - \half \log (1-t_1),
\eeq
and (\ref{eq:mainresult}) we have
\beq
\frac{\partial\overline{E_0}}{\partial t_1} = - \frac{1}{\epsilon^2}
\pv + \frac{1}{2(1-t_1)} .
\eeq
Rewriting the derivative on the left-hand side in terms of $\{
\gpbar \}$ gives
\beq
\sum_{p=2}^{\infty} p \ \gpbar  \ \frac{\partial \egbar}{\partial\gpbar} =
- \frac{(1-t_1)}{\epsilon^2} \pv + \half .
\eeq
However, $\punct_{_V}$ can be written as a power series; using equation
(\ref{eq:partseries}), after summing over $n_0$ and differentiating
with respect to $t_0$, we have
\beq
\pv = \sum_{n_1=0}^{\infty} \ \sum_{n_2=0}^{\infty} \cdots 
\left( \prod_{j=1}^\infty \frac{t_j^{n_j}}{n_j!} \right)
\frac{ {t_0}^{
2+\sum\limits_{k=2}^{\infty}(k-1)n_k} }{ 
\left( 2+\sum\limits_{k=2}^{\infty}(k-1)n_k \right)!} \left(
\sum_{k=1}^\infty k n_k \right) ! \ .
\eeq
But,
\beq
\sum_{n_1=0}^\infty \frac{t_1^{n_1}}{n_1!}  \left(\sum_{k=1}^{\infty}
k n_k \right) ! = \left( \sum_{k=2}^\infty k n_k \right)! \ \left( 1 -
t_1 \right)^{-1- \sum\limits_{k=2}^\infty k n_k}
\eeq
and so,
\beq
\frac{(1-t_1)}{\epsilon^2} \pv = \sum_{n_2=0}^{\infty} \ 
\sum_{n_3=0}^{\infty} \cdots \prod_{j=2}^\infty
\left[\frac{1}{n_j!} \left( \frac{t_j \epsilon^{j-1}}{(1-t_1)^j}
\right)^{n_j} \right] \frac{\left(\sum\limits_{k=2}^\infty k n_k
\right)!}{\left(2 + \sum\limits_{k=2}^\infty (k-1) n_k \right) !}.
\eeq
Thus we have
\beq
\sum_{p=2}^\infty p \ \gpbar \ \frac{\partial
\overline{E_0}}{\partial\gpbar} = - \sum_{\{n_i\}'}
\frac{\left(\sum\limits_{k=2}^\infty k n_k
\right)!}{\left(2 + \sum\limits_{k=2}^\infty (k-1) n_k \right) !}
\ \prod_{p=2}^\infty \left[ \frac{1}{n_p!} \left( - \frac{\gpbar
(2p)!}{p!(p-1)!}  \right)^{n_p}
\right] ,
\eeq
where we are summing each $n_i$ (for $i=2,3,4,\cdots$) from $0$ to
$\infty$, but are excluding the case for which all the $n_i$ are
zero. Integrating this equation yields
\beq
\egbar = - \sum_{\{n_i\}'}
\frac{\left( \left(\sum\limits_{k=2}^\infty k n_k \right) -1
\right)!}{\left(2 + \sum\limits_{k=2}^\infty (k-1) n_k \right) !}
\ \prod_{p=2}^\infty \left[ \frac{1}{n_p!} \left( - \frac{\gpbar
(2p)!}{p!(p-1)!}  \right)^{n_p} \right] .
\eeq
Note that for the special case in which only $\overline{g_2} \neq 0$,
this reduces to
\beq
\overline{E_0}(\overline{g_2}) = - \sum_{n=1}^\infty
\frac{(2n-1)!}{(n+2)!n!} \ \left(-12 \ \overline{g_2} \right)^n,
\eeq
which is the well-known formula counting planar $\Phi^4$
graphs~\cite{BIPZ}. Using (\ref{eq:barnobar}) gives
\beq
E_0(g) = - \sum_{\{n_i\}'}
\frac{\left( \left(\sum\limits_{k=1}^\infty k n_k \right) -1
\right)!}{\left(2 + \sum\limits_{k=2}^\infty (k-1) n_k \right) !}
\ \prod_{p=1}^\infty \left[ \frac{1}{n_p!} \left( - \frac{g_p
(2p)!}{p!(p-1)!}  \right)^{n_p} \right] ,
\eeq
where the sum is now over all values of $n_1,n_2,\cdots$ from $0$ to
$\infty$, with them not all zero. This formula completely solves the
problem of counting planar graphs, whose vertices all have even
coordination numbers.

\subsection{Matrix Model Loop Correlators}
\label{sec:mmloop}
Let us define loop correlators for the reduced Hermitian matrix model
(\ref{eq:zred}) as follows,
\beq
W(p_1) = \frac{1}{N} \sum_{k_1=0}^\infty \frac{ \llang \tr \Phi^{2k_1}
\rrang}{p_1^{2k_1 +1}}
\eeq
and for $s \geq 2$,
\beq
W(p_1, \cdots , p_s) = N^{s-2} \sum_{k_1,\cdots,k_s =1}^\infty
\frac{\llang \tr \Phi^{2k_1} \cdots \tr \Phi^{2k_s} \rrang_{\rm
conn.}}{ p_1^{2k_1 +1} \cdots p_s^{2k_s +1}} .
\eeq
In fact, we will only be interested in the genus zero part of
these functions, which we will denote $W_0(p_1, \cdots, p_s)$; these are
gained by taking the limit $N\to \infty$.

Using equation (\ref{eq:oneptrelate}) we can write $W_0(p_1)$ in terms of
expectation values for the corresponding topological model,
\beq
W_0(p_1) = \frac{1}{p_1} + \sum_{k_1=1}^\infty
\frac{(2k_1)!}{k_1!(k_1-1)!} \ \frac{\llang \sigma_{k_1-1}\rrang_{_V}}{
p^{2k_1+1} \epsilon^{k_1+1}}.
\eeq
This gives using (\ref{eq:onepoint}),
\beq
\label{eq:wonecont}
W_0(p_1) = \frac{1}{p_1} + \left(\frac{2}{p_1^3 \epsilon^2}\right) \pint
\left(1- \frac{z}{z_1} \right)^{-\frac{3}{2}} \left(z - V \right) \zvlog,
\eeq
where we have defined  $z_1= \quart \epsilon  p_1^2$.
Integrating by parts gives
\beq
W_0(p_1) = - \left( \frac{1}{p_1 \epsilon} \right) \pint \left(1-
\frac{z}{z_1} \right)^{-\half} \left(1-V'\right) \zvlog ,
\eeq
where $V'(z) \equiv \frac{dV}{dz}$.
For example, in the case of the Gaussian model, $U(\Phi)=\half \Phi^2$, we have
that $V=\epsilon$ and integrating gives
\beq
W_0(p_1) = \left( \frac{1}{p_1 \epsilon} \right) \int_0^u dx \
\left(1- \frac{x}{z_1} \right)^{-\half} = \frac{p_1}{2}
\left[1-\sqrt{1-\frac{4}{p_1^2}} \ \right] 
\eeq
in agreement with the known result~\cite{JanReview}.

The general loop correlator is given by
\beq
W_0(p_1,\cdots,p_s) = \frac{d}{dU(p_s)} \cdots \frac{d}{dU(p_2)} W_0(p_1),
\eeq
where
\beq
 \frac{d}{dU(p)}= - \sum_{k=1}^\infty \frac{1}{p^{2k+1}}
 \frac{\partial}{\partial g_k} = \sum_{k=1}^\infty \frac{1}{p^{2k+1}}
 \frac{(2k)!}{k!(k-1)!} \ \frac{1}{\epsilon^{k-1}}
 \frac{\partial}{\partial t_k}.
\eeq
Acting with this operator on the contour integral (\ref{eq:wonecont})
gives for $s\geq 2$,
\beq
\label{eq:wtwocont}
W_0(p_1,\cdots,p_s)= - \frac{1}{\epsilon^2} \left( \prod_{i=1}^s
\frac{2}{p_i^3} \right) \pint \left[ \prod_{i=1}^s \left(1-
\frac{z}{z_i} \right)^{-\frac{3}{2}} \right] z^{s-1} \left(\dde
\right)^{s-2} \zvlog ,
\eeq
where
\beq
z_i= \quart \epsilon p_i^2.
\eeq
Note that $W_0(p_1,\cdots,p_s)$ does not depend on the value of
$\epsilon$; this parameter in the mapping between the matrix model
and topological gravity can be freely chosen. It is convenient to
choose $\epsilon=4$. Then for $s=2$ the contour integral can be
converted to a real integral and calculated giving
\beq
W_0(p_1,p_2) = \half \frac{1}{(p_1^2-p_2^2)^2} \left[ p_1^2
\sqrt{\frac{p_2^2-u}{p_1^2-u}} + p_2^2 \sqrt{\frac{p_1^2-u}{p_2^2-u}} -
2 p_1 p_2 \right] ,
\eeq
where $u$ is the solution of $u=V(u)$ for $\epsilon=4$ and we have
assumed that $p_1, p_2>0$.

For $s \geq 3$,
\beq
W_0(p_1,\cdots,p_s)= \frac{1}{\epsilon^2} \left( \prod_{i=1}^s
\frac{2}{p_i^3} \right) \pint \left[ \prod_{i=1}^s \left(1-\frac{z}{z_i}
\right)^{-\frac{3}{2}} \right] z^{s-1} \left(\dde \right)^{s-3} 
\frac{1}{z-V} ,
\eeq
which is being integrated around the pole at $z=u$. This can be easily
evaluated to give
\beq
W_0(p_1,\cdots,p_s)= \left( \frac{1}{M} \frac{\partial}{\partial u}
\right)^{s-3} \left( \frac{1}{u M} \prod_{i=1}^s
\frac{u}{2\left(p_i^2 -u \right)^{\frac{3}{2}} } \right) ,
\eeq
where $M(u)=\quart \left( 1 - V'(u)\right)$ and again $\epsilon=4$.
These formulae, which
were calculated for the reduced Hermitian matrix model, are identical
in form to those for the loop correlators of the complex matrix
model~\cite{AmbJurMak90}, showing the close connection between these two
models at genus zero.

Next we take inverse Laplace transforms of $W_0(p_1,\cdots,p_s)$ with
respect to each of the variables $p_i^2$ to get a new set of loop operators,
\beq
w_0(l_1) = \frac{1}{N} \sum_{n_1=0}^\infty
\frac{l_1^{n_1-\half}}{\Gamma (n_1+\half )} 
\llang \tr \Phi^{2 n_1} \rrang
\eeq
and for $s \geq 2$,
\beq
w_0(l_1,\cdots,l_s)= N^{s-2} \sum_{n_1,\cdots,n_s=1}^\infty
\frac{l_1^{n_1-\half}}{\Gamma (n_1+\half )} \cdots
\frac{l_s^{n_s-\half}}{\Gamma (n_s+\half )} 
\llang \tr \Phi^{2 n_1} \cdots \tr \Phi^{2 n_s}\rrang_{\rm conn.} .
\eeq
Taking inverse Laplace transforms of (\ref{eq:wtwocont}) gives
\beq
w_0(l_1,\cdots,l_s)= - 2^{s-4} \pint \left[ \prod_{i=1}^s
\frac{\sqrt{l_i} e^{z l_i}}{\Gamma (\frac{3}{2} )} \right]
z^{s-1} \left. \left(\dde \right)^{s-2} \zvlog
\right\vert_{\epsilon=4} .
\eeq
Hence for $s \geq 2$,
\beq
\label{eq:wmattop}
w_0(l_1,\cdots,l_s)= \frac{4^{s-2}}{\left[\Gamma(\half)\right]^s}
\sqrt{l_1 \cdots l_s} \left(\frac{\partial}{\partial L} \right)^{s-1}
\llang W(l_1) \cdots W(l_s) \rrang_{\epsilon=4} ,
\eeq
where $L=\sum\limits_{i=1}^s l_i$ as usual and the expectation value
of loop operators on the right-hand side is evaluated for the
potential $V(z)$ with $\epsilon=4$.
The one loop correlator is
\beq
w_0(l_1) = \frac{\sqrt{l_1}}{\Gamma(\half)} \left[ \frac{1}{l_1} +
\quart  \llang W(l_1) \rrang_{\epsilon=4} \right].
\eeq
These equations relate the matrix model loop correlation functions to the
topological gravity loop operators which were studied in
section~\ref{sec:loop}. Using the results in that section one can
rewrite (\ref{eq:wmattop}) as 
\beq
w_0(l_1,\cdots,l_s)= \left. \frac{4^{s-2}}{\left[\Gamma(\half)\right]^s}
\sqrt{l_1 \cdots l_s} \left(\frac{\partial}{\partial L} \right)^{s-1}
\left(\dde \right)^{s-2} \left[ \frac{e^{L u} -1}{L} \right] 
\right\vert_{\epsilon=4} .
\eeq
It is worth noting that in these formulae $u$ is proportional to
$\epsilon$. For example, consider the case of pure gravity, $U(\Phi) = \half
\Phi^2 + g_2 \Phi^4$. If we calculate the corresponding $V(z)$ then
the equation $u=V(u)$ gives
\beq
u = \frac{\epsilon}{24 g_2} \left[ \sqrt{1+48 g_2} -1 \right].
\eeq
This can be inserted into the previous formula to give the multi-loop
correlator for pure gravity in the reduced Hermitian matrix model.

\section{Multi-Critical Hermitian Matrix Models}
\label{sec:hermmulti}
\subsection{Definition of the Matrix Model}

In this section we consider the multi-critical Hermitian matrix models
and compare them with the corresponding models in topological
gravity. The matrix model will be defined as
\beq
\label{eq:hermdef}
Z_N = \int {\cal D} \Phi \ \exp \left[ - \frac{N}{\epsilon}
\tr U(\Phi) \right],
\eeq
where the potential is
\beq
U(\Phi) = \half \Phi^2 - \sum_{n=1}^\infty \overline{\mu_n} U_n(\Phi)
\eeq
and the potential $U_n(\Phi)$ corresponds to that for the $n$-th
multi-critical model,
\begin{eqnarray}
U_1(\Phi) &=& \half \Phi^2 , \\
U_2(\Phi) &=&  \Phi^2 - \frac{1}{12} \Phi^4, \\
& \vdots & \\
U_n(\Phi) &=& \sum_{p=1}^n (-1)^{p-1} \Phi^{2p} \frac{n!
(p-1)!}{(n-p)!(2p)!} \ ;
\end{eqnarray}
these potentials were derived in reference~\cite{GroMig}.

The relationship between this matrix model and topological gravity is
basically the same as in section~\ref{sec:rescale}, except that we
have rescaled $\Phi$ by a factor of $\epsilon^{-\half}$; that is, if 
for the given $U(\Phi)$, the variables $\{g_p\}$ are defined by
equation~(\ref{eq:upot}),
then the parameters in the topological model are given by
\beq
\label{eq:gptotp}
t_p = - \frac{g_p (2p)!}{p!(p-1)!} \mgap {\rm for \ } p \geq 1 .
\eeq
The potential
for the topological model corresponding to the above matrix model is
\beq
\label{eq:potmultipert}
V(z) = \epsilon - \sum_{n=1}^\infty \overline{\mu_n} \left[ (1-z)^n -1 \right] 
\eeq
and the $k$-th multi-critical model is gained by choosing $\epsilon
\neq 0$, $\overline{\mu_1}=1$ and $\overline{\mu_k}=-1$; that is,
\beq
\label{eq:vkth}
V(z)= \epsilon + z + (1-z)^k -1 \equiv V_k,
\eeq
which we have denoted $V_k$.
Note that this is different from the potential usually chosen in
topological gravity for the $k$-th multi-critical model, namely $V=
\epsilon +z - z^k$; this potential, which has $t_1=1$, would give
$g_1=-\half$ and hence $U(\Phi)=g_k \Phi^{2k}$. Lacking a $\Phi^2$
term in the exponential, this matrix model 
does not correspond to the usual picture
of surfaces made by gluing together polygons.
This is the source of much of the confusion that
has arisen when comparing the multi-critical matrix and topological
models.

\subsection{Correlation Functions in the Multi-Critical Topological Model}
We begin by calculating the correlation functions for the topological
model with the potential $V_k$ given in equation (\ref{eq:vkth}). Then in
the next section we consider the connection between these correlation
functions and those in the matrix model.

For the $k$-th multi-critical model, $u$ is given by $u=V_k(u)$ and hence,
\beq
u= 1 - (1- \epsilon)^{\ok} = 1 - (\deps)^{\ok} ,
\eeq
where we have defined $\deps \equiv 1 - \epsilon$.

The partition function for the $k$-th multi-critical model is given by
substituting $V_k(z)$ into (\ref{eq:contour}),
\beq
\part_{_{V_k}} = \half \int_0^u dx \ \left( \deps - (1-x)^k \right)^2 = 
- \half \int_1^{\depsok} \back dy \ ( \deps - y^k )^2 ,
\eeq
where we have changed variables to $y=1-x$.
Thus dropping the regular terms,
\beq
\part_{_{V_k}} \sim  \frac{- k^2}{(2k+1)(k+1)} (\deps)^{2+\ok} ;
\eeq
this scales in the same fashion as (\ref{eq:partscale}), which was
calculated for the $k$-th multi-critical model using the simpler
potential. 

Next we define a set of operators, $\{O_l\}$,
\beq
\label{eq:defol}
O_l = - \sum_{n=0}^l (-1)^n \left( {l \atop n} \right) \sigma_n ,
\eeq
which are chosen so that they scale in the correct fashion. The
operator $O_l$ gives a factor of $-(1-z)^l$ in the contour integral
and hence using (\ref{eq:onepoint}),
\beq
\llang O_l \rrang_{_{V_k}}
 = - \int_1^{\depsok} \back dy \ y^l (\deps -
y^k ) \sim   \frac{-k}{(l+k+1)(l+1)} (\deps)^{1+(l+1)/k},
\eeq
where we have dropped the regular terms in the last expression.

Using (\ref{eq:genfunc}), the general case (for $s \geq 2$) is given
by
\begin{eqnarray}
\label{eq:genocorr}
\llang \prod_{i=1}^s O_{l_i} \rrang_{_{\back V_k}}
 &=& - \left( - \dde
\right)^{s-2} \int_1^{\depsok} dy \ y^L \\
& \sim & - \left(\frac{\partial}{\partial \deps} \right)^{s-2}
\left[ \frac{1}{(L+1)} (\deps)^{(L+1)/k} \right] ,
\end{eqnarray}
where $L=\sum\limits_{i=1}^s l_i$. 

Thus, up to regular terms, 
the scaling with respect to
$\deps$ of correlation functions of
$O_l$ for the potential, $V_k=\epsilon  + z + (1-z)^k -1$,
is the same as the scaling with respect to $\epsilon$ of
correlation functions of $\sigma_l$
using the simpler potential, $V= \epsilon + z - z^k$; these were
calculated in section~\ref{sec:multitop}.
Note that some papers confuse $\sigma_l$ with $O_l$, however they are
in fact related by equation (\ref{eq:defol}).

\subsection{Correlation Functions in the Multi-Critical Matrix Model}
In reference~\cite{GroMig}, Gross and Migdal study correlation
functions for the $k$-th multi-critical Hermitian matrix model at the
spherical level. The
correlators that they calculate scale in exactly the same fashion as
those we derived in the previous section for topological gravity. It
will be instructive to examine the connection between the two
calculations. 

In their paper, the free energy of the matrix model is $-\log Z_N$;
the partition function for this model can be written 
as in equation (\ref{eq:hermdef}). 
After dropping a number of terms,
which they claim do not contribute to the critical behaviour,
the free energy is given by $F$, where 
\beq
\frac{\partial^2 F}{\partial t^2} = f(t) 
\eeq
and $f(t,\{\mu_n\})$ is the relevant solution of 
\beq
t=f^k - \sum_{n=0}^{\infty} \mu_n f^n .
\eeq
We will take the last two equations as being the definition of $F$
and then later consider the connection between $F$ and the free energy
of the matrix model. The correlation functions are defined as
\beq
\label{eq:corrf}
- \left. \frac{\partial}{\partial \mu_{l_1}} \cdots
\frac{\partial}{\partial \mu_{l_s}} F(t,\{\mu_n\})
\right\vert_{\{\mu_n=0\}} .
\eeq
It should be noted that they introduced various powers of $N$ in their
definitions of the scaling variables: $t$, $f(t)$ and so on. However
these factors of $N$ do not affect the following arguments and can
safely be ignored.

To make the connection with topological gravity we consider the
potential,
\beq
\label{eq:grosspot}
V(z)= -t + z + (1-z)^k - \sum_{n=0}^\infty \mu_n (1-z)^n .
\eeq
This is just equation (\ref{eq:potmultipert}) with the change of variables,
\beq
\overline{\mu_n} = \mu_n + \delta_{n,1} - \delta_{n,k}
\eeq
and
\beq
\epsilon = 1 - t - \sum_{n=0}^\infty \mu_n .
\eeq
The potential gives the $k$-th multi-critical model perturbed by terms
corresponding to the operators $O_n$. Then the equation $u=V(u)$ can
be written,
\beq
t= (1-u)^k - \sum_{n=0}^\infty \mu_n (1-u)^n .
\eeq
Thus we see that $f=1-u$, where $u=\llang P P \rrang$ in
this topological model as was shown in section~\ref{sec:multitop}. 
However $\dde = - \frac{\partial}{\partial t}$,
hence,
\beq
\frac{\partial^2 F}{\partial t^2}= f = 1- \langle PP \rangle = 1 -
\frac{\partial^2}{\partial t^2} \part .
\eeq
Integrating, we have that $F= - \part$, up to some regular terms which
we will drop. Thus $F$, which is supposed to scale in the
same way as the free energy of the matrix model, is in fact
proportional to the partition function of the topological model
defined in (\ref{eq:grosspot}), up to some regular terms. Noting that
$t= \deps$, when $\mu_n=0$ for all $n$, then,
\beq
- \left. \frac{\partial}{\partial \mu_{l_1}} \cdots
\frac{\partial}{\partial \mu_{l_s}} F
\right\vert_{\{\mu_n=0\}} = \llang O_{l_1} \cdots O_{l_s}
\rrang_{_{V_k}} ,
\eeq
since deriving with respect to $\mu_l$ just pulls down an $O_l$
operator. Note also that the correlation function on the right-hand side is
calculated in topological gravity for the $k$-th multi-critical model
(we calculated this in the previous section),
it is {\it not} a matrix model
correlator. Thus Gross and Migdal have been calculating
topological correlation functions in their paper. The crucial question
to answer is the following one: 
``If $F$ in equation (\ref{eq:corrf}) is replaced by the
free energy of the matrix model, do the correlation functions still
scale in the same fashion?''
The answer is ``Yes, provided that $F$ is replaced by $\epsilon^2
E_0$.'' The factor of $\epsilon^2$ in front of the free energy is
necessary, as without it there are extra terms in the correlation
functions, which would change the scaling behaviour if any of the
$l_i$ were greater than $k$.

The free energy of the matrix model, $E_0$, is defined as in equation
(\ref{eq:freeenergy}). We wish to calculate the correlation functions,
which are derivatives with respect to $\mu_{l_i}$ of $-\epsilon^2
E_0$, evaluated
at the $k$-th multi-critical model (that is, $\mu_n=0$ for all $n$). 
However we know from (\ref{eq:defol}) that
\beq
\label{eq:ddmul}
\frac{\partial}{\partial \mu_{l}} 
= \sum_{n=0}^l (-1)^{n+1} \left( {l
\atop n} \right) \frac{\partial}{\partial t_n} .
\eeq
Using equations (\ref{eq:mainresult}) and (\ref{eq:dedeps})
one can calculate the effect of this operator on $\epsilon^2 E_0$; note that
these equations were left unchanged under the rescaling of $\Phi$ by
$\epsilon^{-\half}$ , which we performed earlier. After some algebra,
we have
\beq
\left.
\frac{\partial}{\partial \mu_l} \left(\epsilon^2 E_0 \right)
 \right\vert_{\{\mu_n=0\}}
= - \epsilon \log k + \int_1^{\depsok} \back dy \ (\deps - y^k) \left[
\frac{y^l}{1-y} - \frac{ky^{k-1}}{1-y^k} \right].
\eeq
The leading singular behaviour is given by
\beq
\label{eq:onepte}
\left.
- \frac{\partial}{\partial \mu_l} \left( \epsilon^2 E_0 \right)
 \right\vert_{\{\mu_n=0\}}
 \sim \frac{-k}{(l+k+1)(l+1)}
(\deps)^{1+(l+1)/k} \sim \llang O_l \rrang_{_{V_k}} ,
\eeq
as expected.

The scaling of $E_0$ can also be calculated from (\ref{eq:onepte}) since,
\beq
\left.
\frac{\partial}{\partial \epsilon} \left(\epsilon^2 E_0 \right)
 \right\vert_{\{\mu_n=0\}}
 = \left. - \frac{\partial}{\partial \mu_0} \left(\epsilon^2 E_0 \right)
 \right\vert_{\{\mu_n=0\}}
\sim \frac{-k}{(k+1)} (\deps)^{1+\ok}
\eeq
and hence as expected,
\beq
\left.
\epsilon^2 E_0  \right\vert_{\{\mu_n=0\}}
 \sim \frac{k^2}{(2k+1)(k+1)} (\deps)^{2+\ok} \sim - \part_{_{V_k}} .
\eeq

In principle one could extend this calculation to multi-point
correlation functions, however the algebra rapidly becomes tedious. In
order to prove that derivatives of $\epsilon^2 E_0$ have the same
leading singular behaviour as derivatives of $F$, it is convenient to
return to reference~\cite{Bessis}. This paper gives an alternative
form of equation (\ref{eq:egbarw}),
\beq
\egbar = - \int_0^1 dx \ (1-x) \log\left( \frac{r_0(x)}{x}\right) ,
\eeq
where $r_0(x)$ is defined by $x=w(r_0)$. This leads to
\beq
E_0(g) = \half \log \epsilon - \frac{1}{\epsilon^2} \int_0^\epsilon dx
\ (\epsilon - x) \log u(x) ,
\eeq
where $u(x)$ is the solution of $V(u(x))=u(x)$ for the potential
(\ref{eq:pot}) with $t_0=x$. Thus,
\beq
\frac{\partial^2}{\partial \epsilon^2} \left(\epsilon^2 E_0 \right) =
\frac{3}{2} + \log \epsilon - \log u(\epsilon) .
\eeq
Dropping the first two terms, which are regular in $\deps$,
we can differentiate with respect to $\mu_{l_1}$,
\beq
\frac{\partial^2}{\partial \epsilon^2} \left[ \frac{\partial}{\partial
\mu_{l_1}} \left(\epsilon^2 E_0 \right) \right] \sim -
\frac{\llang P P O_{l_1} \rrang}{\llang P P
\rrang} \sim - \llang P P O_{l_1} \rrang,
\eeq
where the last expression uses the fact that, $u(\epsilon) = \llang P P
\rrang = 1 - \depsok$ and we are keeping only the leading
non-analytic term. Integrating gives, up to regular and sub-leading terms,
\beq
\left.
- \frac{\partial}{\partial \mu_{l_1}} \left( \epsilon^2 E_0 \right)
 \right\vert_{\{\mu_n=0\}}
\sim \llang O_{l_1} \rrang_{_{V_k}} .
\eeq
This can easily be generalized to give
\beq
\label{eq:freescale}
\left.
- \frac{\partial}{\partial \mu_{l_1}} \cdots
  \frac{\partial}{\partial \mu_{l_s}}
\left( \epsilon^2 E_0 \right)  \right\vert_{\{\mu_n=0\}}
\sim \llang O_{l_1} \cdots O_{l_s}\rrang_{_{V_k}} .
\eeq
Thus as claimed earlier, the correlation functions for the reduced
Hermitian matrix model scale in the same fashion as the topological
correlation functions, for the $k$-th multi-critical model. 

It should perhaps be noted at this point that if
one rescales all the variables by various powers of $N$ as in
reference~\cite{GroMig}, then in the spherical limit $N \to \infty$
all the sub-leading terms in (\ref{eq:freescale}) vanish (this
rescaling works provided we have dropped the regular terms).
However, there is a simpler way of removing the sub-leading
terms. From equation (\ref{eq:mainresult}) we have that
\beq
- \left( \frac{\partial}{\partial \mu_l} - \frac{\partial}{\partial \mu_{l+1}}
\right) \left(\epsilon^2 E_0 \right) = \llang O_l
\rrang_{_V} ,
\eeq
which is true for a generic potential $V(z)$. Differentiating
repeatedly and evaluating for $V_k$ gives
\beq
\label{eq:corrnosub}
- \left. \left( \frac{\partial}{\partial \mu_{l_1}} -
 \frac{\partial}{\partial \mu_{l_1+1}} \right) 
\frac{\partial}{\partial \mu_{l_2}} \cdots
\frac{\partial}{\partial \mu_{l_s}}
\left(\epsilon^2 E_0 \right) \right\vert_{\{\mu_n=0\}}
 = \llang O_{l_1} \cdots O_{l_s}
\rrang_{_{V_k}}.
\eeq
This equation is exact and there are no sub-leading terms.

Using (\ref{eq:ddmul}) and (\ref{eq:gptotp}), one can write
$\frac{\partial}{\partial \mu_l}$ in terms of derivatives with respect
to the variables $\epsilon$ and $g_p$,
\beq
\label{eq:mulop}
\frac{\partial}{\partial \mu_l} = \sum_{p=1}^l (-1)^p
\frac{l!(p-1)!}{(l-p)!(2p)!} \frac{\partial}{\partial g_p} - \dde .
\eeq
In principle one can use this to operate on $\epsilon^2 E_0$ and hence
rewrite the correlation functions (\ref{eq:corrnosub}) as matrix model
expectation values. The summation in (\ref{eq:mulop}) brings
down a term proportional to
$\tr U_l(\Phi)$, but the derivative with respect to $\epsilon$
is more complicated and gives a number of terms including one proportional
to $\tr U(\Phi)$ in the expectation value. As a result the multi-point
correlators are not particularly simple. Calculating the one-point
correlation function is however straightforward,
\beq
\llang O_l \rrang_{_V} =
- \left( \frac{\partial}{\partial \mu_l} - 
\frac{\partial}{\partial \mu_{l+1}}
\right) \left(\epsilon^2 E_0 \right) = \lim_{N \to \infty}
\frac{\epsilon}{N} \llang \tr \left[ U_l(\Phi) - U_{l+1}(\Phi)
\right] \rrang.
\eeq
In fact, defining
\beq
\utw_l \equiv \tr \left[ U_l(\Phi) - U_{l+1}(\Phi) \right] ,
\eeq
\beq
\ubar_l \equiv \tr \left[ U_l(\Phi) -  \frac{1}{\epsilon} U(\Phi) \right] 
\eeq
and
\beq
\frac{\partial}{\partial \widetilde{\mu_l}} \equiv
\frac{\partial}{\partial  \mu_l} -
\frac{\partial}{\partial  \mu_{l+1}},
\eeq
we have for the first few multi-point correlation functions in the
spherical limit $N \to \infty$,
\beq
\llang O_{l_1} O_{l_2} \rrang_{_V} = -
\frac{\partial}{\partial \widetilde{\mu_{l_1}}}
\frac{\partial}{\partial \mu_{l_2}}  \left(\epsilon^2 E_0 \right) =
\llang \utw_{l_1} \ubar_{l_2} \rrang_{\rm conn.} -
\frac{1}{N}  \llang \utw_{l_1} \rrang ,
\eeq
\beq
\llang O_{l_1} O_{l_2} O_{l_3} \rrang_{_V} = -
\frac{\partial}{\partial \widetilde{\mu_{l_1}}}
\frac{\partial}{\partial \mu_{l_2}}
\frac{\partial}{\partial \mu_{l_3}} \left(\epsilon^2 E_0 \right) =
\left(\frac{N}{\epsilon}\right)  \llang \utw_{l_1} \ubar_{l_2}
\ubar_{l_3} \rrang_{\rm conn.}
\eeq
and
\begin{eqnarray}
\nonumber
\llang O_{l_1} O_{l_2} O_{l_3} O_{l_4} \rrang_{_V} &=& 
\left(\frac{N}{\epsilon}\right)^2 \biggl[
  \llang \utw_{l_1} \ubar_{l_2}
\ubar_{l_3} \ubar_{l_4} \rrang_{\rm conn.} \\
& & + \frac{1}{N} \llang
\utw_{l_1} \left( \ubar_{l_2} \ubar_{l_3} +  \ubar_{l_3} \ubar_{l_4} +
\ubar_{l_4} \ubar_{l_2} \right) \rrang_{\rm conn.} \biggr] .
\end{eqnarray}

The correlation functions in topological gravity and the Hermitian
matrix model can be related more simply at the cost of reintroducing
sub-leading terms. If we define,
\beq
\widetilde{O}_l = O_l - O_{l+1} ,
\eeq
then in the spherical limit we have
\beq
\llang O_{l_1} \rrang_{_V} = \left(\frac{\epsilon}{N}\right)
\llang \utw_{l_1} \rrang 
\eeq
and in general,
\beq
\llang O_{l_1} \prod_{i=2}^s \widetilde{O}_{l_i} \rrang_{_{\back V}} =
\left(\frac{N}{\epsilon}\right)^{s-2} 
\llang \prod_{i=1}^s \utw_{l_i} \rrang_{\rm
conn.} .
\eeq
This completes our study of the connection between the correlation
functions in the reduced Hermitian matrix model and those in
topological gravity.

\section{Conclusion}
\label{sec:conc}

We have argued that it is possible in 2d topological gravity to
consider observables depending on geodesic distances, such as the
volume of space-time and the length of the boundaries. We then
constructed a detailed mapping between topological gravity and the
reduced Hermitian matrix model, both for operators and coupling
constants.  In particular, the operators in topological gravity were
associated with very specific matrix model operators even before the
continuum limit of the matrix model was taken.  As a spin-off of this
identification we could use results from topological gravity to
provide the complete solution of the counting problem for planar
graphs whose vertices have all even coordination numbers. It is an
interesting problem to consider whether these counting results can be
extended to arbitrary genus. Further, the simple and exact relation
between a set of planar graphs and the partition function of
topological gravity hints that there may be a simpler interpretation
of the intersection indices on the moduli spaces of Riemann surfaces
with punctures. However, a substantiation of  such a claim requires first a
generalization of the results of this paper to surfaces of arbitrary
genus.  An alternative representation of the operators $\sg_n$ might
allow us to understand one remaining puzzle in topological gravity, namely 
the transition from local operators $\sg_n$, which create only
punctures in the Riemann surfaces, to operators which create
macroscopic boundaries.  Using matrix models such a transition is
conceptionally simple since $\tr \Phi^{2n}$ will create a macroscopic
boundary, if $n \to \infty$ as the square root of the (discretized) volume. 
Since we have 
shown that $\la \sg_{n-1} \ra $ is closely related to $\tr \Phi^{2n}$,
the possibility of a geometric interpretation of  $\la \sg_{n-1} \ra$ 
for $n \to \infty$ should exist. We hope to be able to return to this 
question in a later publication.

\subsection*{Acknowledgements}
MGH would like to acknowledge the support of the European Union
through their TMR Programme. MW acknowledges the support of the Danish
Natural Science Research Council (grant 11-0801).


\end{document}